# Deploying Scientific AI Networks at Petaflop Scale on Secure Large Scale HPC Production Systems with Containers.

David Brayford
LRZ
Munich Germany
brayford@lrz.de

Sofia Vallecorsa
CERN
Geneva Switzerland
sofia.vallecorsa@cern.ch

## ABSTRACT

There is an ever-increasing need for computational power to train complex artificial intelligence (AI) & machine learning (ML) models to tackle large scientific problems. High performance computing (HPC) resources are required to efficiently compute and scale complex models across tens of thousands of compute nodes. In this paper, we discuss the issues associated with the deployment of machine learning frameworks on large scale secure HPC systems and how we successfully deployed a standard machine learning framework on a secure large scale HPC production system, to train a complex three-dimensional convolutional GAN (3DGAN), with petaflop performance. 3DGAN is an example from the high energy physics domain, designed to simulate the energy pattern produced by showers of secondary particles inside a particle detector on various HPC systems.

## 1 Introduction

The convergence of HPC and AI is becoming increasingly important in analyzing the massive amounts of data generated by numerical simulations and scientific experiments in fields such as physics, computational chemistry and weather forecasting. In addition, detecting diseases and personalized medicine are some of the examples for very important applications in medical research, which can improve our daily lives. Financial security is also one of the applications, where the use of machine learning enhances financial institutions' ability to detect credit card fraud, money laundering and other criminal activities.

Present-day, processor technology has made great progress up to the point where even general purposes CPUs can execute deep learning (DL) workloads. However, the size of the neural networks, are getting larger along with the amount of data required to solve these complex problems and are representing a new challenge for the data science community.

Among the most cost-effective solutions is to use pre-existing HPC infrastructures to train these complex models. They provide the possibility of scaling out to more compute nodes and parallelize the workloads in order to speed-up the training.

As a result, scalability of the computational resources is the key to solve extremely large problems involving complex neural networks and to take advantage of the massive amount of computational power available on large scale production HPC systems.

Enabling data scientists to efficiently use such systems is challenging, due to the scarcity of information, technical knowledge and software, which can accommodate both AI and HPC needs. For example, scientists have started to use deep learning techniques to analyze the massive amounts of data generated during their experiments. This type of analysis cannot be achieved on traditional systems used by data scientists and requires HPC resources to cope with the amount of computation, data storage and data transfer required.

However, the deployment of ML frameworks and workflows on HPC systems is not straightforward and requires a lot of time and effort by the data scientists and HPC centers to build and maintain the different frameworks. For example, it is not always possible to install specific ML software versions, required by the end users, or modify the HPC system software, for instance changing the version of glibc on the system.

To enable data scientists to take advantage of the massive amount of compute available on large HPC systems we need to find a mechanism that enables them to deploy their workflows and software in a way that is simple and does not require a lot of modification to their code and workflow, but also respects the existing HPC system environment, workflows and security policies. To achieve this, we employ the use of Charliecloud a secure HPC container for secure HPC systems developed at Los Alamos National Laboratory (LANL).

The purpose of this paper is to highlight the work done by the Leibniz Supercomputing Center (LRZ) enabling the transition of AI and ML workloads from the laptop to the supercomputer, without compromising security on the HPC systems. In addition, we highlight the performance improvements for the distributed training times of the 3DGAN model on SuperMUC-NG (SNG) and how we achieved almost linear scaling to petascale.

## 2  Related Work

In this section we introduce some of the related technologies, which are commonly used to deploy AI and ML software on computer systems.

### 2.1  AI at Extreme Scale

Scientists have started to explore the deployment of AI technologies on some of the world's largest supercomputers in order to address some of the open challenges that they are facing such as climate science.

The ClimateNet project from Lawrence Berkeley National Laboratory (LBNL) is one example. It employs deep learning methods to identify important weather and climate patterns via expert-labeled, community-sourced open datasets and architectures [1, 2]. One component of the project involves the classification of extreme weather in climate simulations [3] on the Cori HPC system at NERSC, which used a 15 TB dataset from climate simulations, resulting in a peak performance of 15.07 petaflops, and sustained performance of 13.27 petaflops.

The 2018 ACM Gordon Bell prize shared by LBNL and Oak Ridge National Laboratory (ORNL) for applying an exascale-class deep learning application to extreme climate data [4]. Where the Tiramisu model had a sustained performance of 176.8 petaflops for FP32 data precision and 492.2 petaflops for FP16 data precision. While the DeepLabv3+ model obtained a sustained performance of 325.8 petaflops for FP32. The FP16 network reaches a peak performance of 1.13 exaflops and a sustained performance of 999.0 petaflops.

### 2.2  Python

Python [5] is an interpreted high-level programming language, which is used extensively in the fields of pre- and post-processing of data, artificial intelligence and specifically machine learning. The language is widely considered accessible for prototyping algorithms, due to the inherent constructs and the availability of optimized scientific and AI & ML specific libraries, compared to traditional programming languages such as Fortran, C or C++. In addition, because Python is an interpreted language, it does not need to be recompiled for different CPU architectures unlike C or C++.

The tradeoff of having portability, flexibility and ease of use is the loss in performance over code developed in C or C++. However, this is somewhat mitigated by using optimized scientific, numerical libraries such as NumPy, SciPy and scikit-learn provided by the vendors.

Another significant issue with using Python on secure HPC systems is the way Python installs software, which often requires an Internet connection. As a result, deploying Python software on a secure HPC system without an Internet connection is problematic.

In addition, a single instance of Python is not suitable for multi-user, multi-framework usage due to the automatic downgrading and upgrading of dependency software packages. Potentially, the automated installation mechanism can change some or all of the shared dependencies to the version required by the newly installed software package, potentially break the previously installed packages.

### 2.3  User Defined Software Stack

With the increasing demand on more flexible execution models from data scientists and researchers, the standard methods to provide software packages on HPC systems becomes unpractical for the more dynamic AI software stacks. User defined software stacks (UDSS) [6], combined with the recently introduced user namespaces in Linux, offer a solution that can be realized in the form of containers without sacrificing security of the HPC cluster.

The UDSS provide a mechanism to deal with problems such as frequency software updates, handling dependencies, and support for different Linux distributions & version.

User namespaces available in Linux allows the execution of privileged operations to be performed inside the container without escalating permissions up to root on the host environment.

#### 2.3.1  Docker

Docker [7] is considered the industry standard container that provides the ability to package and run an application in an isolated environment.

However, Docker was not designed for use in secure environments and has significant security issues that enables the user inside the Docker container to have privileged (root) access on the host systems, making it unsuitable for HPC systems. In addition, Docker uses cgroups to isolate containers, which can conflict with the Slurm scheduler, which also uses cgroups to allocate resources to batch jobs and enforce limits.

### 2.3.2 Singularity

Singularity [8] was originally developed at LBL to be a containerization solution for HPC systems and supports several HPC components such as resource managers, job schedulers and contains built in MPI features. That enables Docker images to be converted into secure container images that can be run in userspace.

Although Singularity has been developed to run in a non-privileged namespace, potential security issues came to light during a security review at LRZ in 2018, which have since been resolved in later versions. As a result of the internal security review and concerns of the system administrators. LRZ current policy is not to allow Singularity on SNG.

### 2.3.3 Shifter

Shifter [9] was developed at NERSC in collaboration with Cray to enable Docker images to be securely executed on an HPC system. Shifter works by converting Docker images to a common format that can then be distributed and launched on HPC systems. Shifter works by enabling users to convert the Docker images to a flattened format, which are directly mounted on the compute nodes using a loopback device.

Shifter appears to be a good choice for conventional HPC batch queuing infrastructure. However, Shifter was developed to work well on the systems at NERSC and it does not appear to work as well on HPC systems at other centers out of the box. In addition, Shifter requires more administrative setup than other HPC container technologies.

### 2.3.4 Charliecloud

Charliecloud [6] was developed at LANL to be a lightweight open source UDSS implementation based on the Linux user namespace for HPC sites with strict security requirements and consists of approximately 1000 lines of code. It employs Docker to build the Charliecloud image, shell scripts to unpack the image to an appropriate location and a C program to activate the image and run user code within the image.

In these secure environments, Charliecloud's distinct advantage is the separation of the build phase from the runtime and the useage of the newly introduced user namespace to enable non-privileged launch of containerized applications. The user namespace is an unprivileged namespace and within the user namespace, all other privileged namespaces are created without the requirement of root privileges, which means that a containerized application can be launched without requiring privileged access to the host system.

Brayford et al. [10] describe a mechanism that employs Charliecloud to deploy TensorFlow and train a complex neural network at scale on a secure HPC system.

### 2.3.5 Podman

Podman [11] is an open source container management tool for developing, managing and deploying containers on Linux systems. For HPC you usually disable network, PID and IPC namespaces. It runs without any additional permissions, just as a normal user process. The only thing necessary are user namespaces. So, it is as secure as any other process running in a user namespace with Linux.

Podman appears to be a good choice for secure containers as it runs entirely in userspace and is supported by RedHat. However, Podman wasn't on the list of container technologies allowed on the HPC systems at LRZ. It wasn't considered for this work, but will be considered for future work once the system administrators have investigated the security implications.

## 2.4 TensorFlow

TensorFlow [12], developed by Google, is a software framework to develop and run Deep Learning based solutions.

TensorFlow is one of the most widely used machine learning frameworks and it has been deployed on a wide variety of hardware platforms to conduct AI/ML research.

TensorFlow can be built from source either in a standard way or with external packages, which aims at improving performances, like the MKL-DNN library. However, this approach would require all the packages to be pre-downloaded in order to have them available at build time. An additional load, that will be there regardless of any external package, is represented by Bazel: a build tool developed by Google that is required for building TensorFlow. In addition, the release cadence of TensorFlow is significantly higher than most traditional HPC software packages. This means that all modifications to the build process need to be redone on a regular basis.

## 2.5 Horovod

Horovod [13] was developed at Uber for TensorFlow, Keras [14] and PyTorch [15]. It uses the message passing interface (MPI) as mechanism for communication to enable distributed training. It uses MPI operations such as Allreduce and Allgather to handle the communication and weight updates. It implements a ring-allreduce algorithm [13] that reduces communication costs. This is different to the standard TensorFlow hierarchical architecture where workers pass parameter updates to centralized parameter servers. Horovod is installed as a separate Python package and requires MPI libraries and header files to be in the path, otherwise it won't compile.

Calling Horovod's API from the model script, enables a standard build of TensorFlow to run distributed training with minimal changes to the source code required to support distributed training with MPI.

## 3 System & Components

### 3.1 SuperMUC-NG

The SuperMUC-NG (SNG) system at the Leibniz Supercomputing Center of the Bavarian Academy of Science, has a peak performance of 26.9 petaflops, which consists of 311,040 Intel Xeon Platinum 8174 (Skylake-SP) CPU cores and 719 terabytes of main memory. The Skylake cores are arranged into eight "thin" and one "fat" island. The "thin" islands consist of 792 nodes with each node contains 48 CPU cores and 96 Gigabytes of main memory. The "fat" island consists of 144 nodes with 48 CPU cores per node and 768 Gigabytes of main memory.

The systems software stack is OpenHPC [16] compliant and the main development environment is Intel Parallel Studio XE & Intel MPI.

For security reasons, SNG has no direct connection to the Internet on the login nodes, and SSH has been disabled on the compute nodes.

### 3.2 Charliecloud Conversion

To be able to deploy the modified Intel optimized TensorFlow Docker image on SNG it first must be converted into a Charliecloud container. Before we can proceed, Docker and Charliecloud must be installed on the developers Linux system. The Docker and Charliecloud conversion commands necessary to convert the Docker image to a Charliecloud image require privileged (root) access.

If we need to install software in the containerized image that requires access to another system, we must do this on a system that can access the required servers. For example, the command 'pip install' will not succeed because the system does not allow access via https.

The method we employed at LRZ was to download the Intel optimized TensorFlow Docker image from the Intel AI Docker Hub [17]. Then modified the containerized OS environment to support distributed training by installing MPI libraries and Horovod via pip. This modified Docker image was then saved and converted to a Charliecloud image and then copied to the secure HPC system.

## 4 Execution on SuperMUC-NG

In this section, we describe the 3DGAN network developed by CERN and how we deploy a Charliecloud container on SNG at LRZ.

### 4.1 The 3D-GAN Network

Calorimeters are important components of High Energy Physics (HEP) detectors. They are segmented in a large number of cells: by recording the energy deposited in each cell, it is possible to reconstruct the energy pattern produced by showers of secondary particles inside the detector volume. Due to their complexity, calorimeters require a lot of computational power to simulate using classical Monte Carlo based approaches. However, by interpreting each cell as a pixel in an image and the amount of deposited energy per cell as the pixel intensity, it is possible to represent the calorimeter output as a three-dimensional image and apply deep learning models, developed for computer vision problems. 3D Generative Adversarial Network (3DGAN) employs these techniques and uses 3D convolutional Generative Adversarial Networks (GANs) [18] to simulate synthetic energy showers, in the same way they will be recorded by next-generation of high granularity calorimeters [19]. In terms of the physics, the initial validation of 3DGAN results display a remarkable agreement with the state-of-the-art Monte Carlo-based simulation [20].

The training dataset is produced in the context of the current design studies of the Linear Collider Detector for the Compact Linear Collider [21]. That consists of an electromagnetic calorimeter modelled as a regular grid of 3D cells. Each entry in the dataset represents the energy depositions within individual calorimeter cells produced by one electron, stored as a 25x25x25 pixel image [22]. Details about the 3DGAN model are described by Goodfellow et.al [18] and Hinton et.al. [23].

Integration with Horovod and the optimization of TensorFlow and Horovod parameters required to improve the training performance on Intel® Xeon® processors is described in [24]. Initial results on the scaling out of the training achieve approximately 94% scaling efficiency on 128 compute nodes. The scaling of the 3DGAN model across multiple nodes was achieved by using a data parallel training approach employing a synchronous stochastic gradient decent, which was implemented using Horovod. Since the scaling happens along the batch size, it is a weak-scaling approach where the batch size per MPI rank is constant. The advantage of data parallel training is the reduction in communication between processes, which is only required during the back-propagation phase and therefore allows for larger training runs. The downside is that with increasing the number of workers, the global batch size increases and the learning rate has to be increased to ensure that the overall number of epochs remains constant. This increased learning rate has a negative impact on the overall training accuracy, which is described by Caminati et.al. [25].

## 4.2 Slurm Single Node

Deep learning networks are typically trained in an interactive manner on a workstation. However, interactive execute mode is not the standard way of running applications on a HPC system, which employs a batch scheduler such as Slurm.

Within a single shared memory node it is possible to employ both MPI and OpenMP [26] to enable the parallel execution of tasks. However, for most cases OpenMP is considered the most appropriate mechanism for single node execution.

## 4.3 Slurm Multi-Node

The standard execution mode on an HPC system involves distributing the computation across several nodes using MPI as the standard method of communication used to enable the distribution of computation and parallel task execution across the nodes on the system.

To enable the distributed 3DGAN training to be executed from within the Charliecloud container on SNG we created a Slurm script. That contains the number of nodes, number of MPI task, number of OpenMP threads and Charliecloud commands required to execute the training. This information is described in more detail on the High Performance AI Github site [27].

## 5 Results

In this section, we discuss the results of training the distributed 3DGAN model from within the Charliecloud container using MPI on SNG thin nodes. The networks were trained at least 4 times and the results presented are the averages of the epochs.

## 5.1 Multi-Node Scaling within a Container

We were able to train the distributed 3DGAN model on SNG using the containerization method described previously. The training of the network was performed using hybrid MPI and OpenMP parallelism. That involved one MPI rank per node and one OpenMP thread per physical CPU core. Table 1 shows the time taken to complete a single epoch. All jobs were submitted to the system via the Slurm batch system.

| Nodes | Training Time(S) per Epoch | Linear Time(S) per Epoch | Scaling Efficiency |
|---|---|---|---|
| 4 | 3806 | 3806 | - |
| 8 | 1910 | 1903 | 99.6% |
| 16 | 1001 | 951.5 | 95.1% |
| 32 | 504 | 475.75 | 94.4% |
| 64 | 253 | 237.87 | 94% |
| 128 | 124 | 118.93 | 95.9% |
| 256 | 61 | 59.46 | 97.5% |
| 512 | 33 | 29.73 | 90.1% |

**Table 1: Shows the time taken in seconds for a single epoch of the CERN 3D-GAN network to complete.**

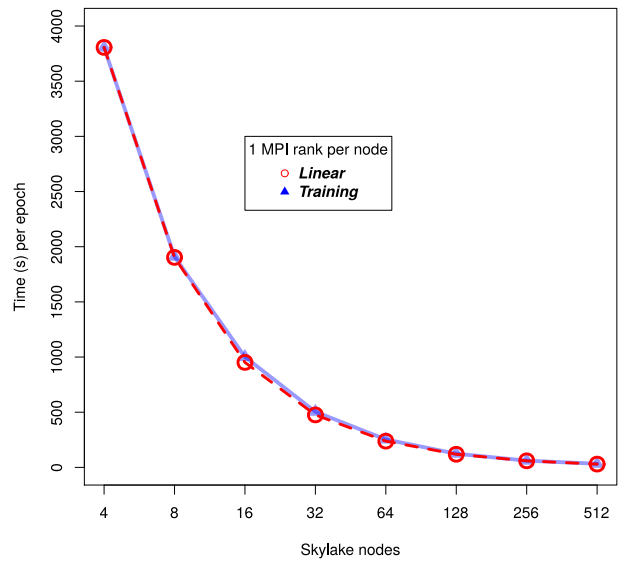

**Figure 1: Scaling plot for the training of the TensorFlow on SNG with 1 MPI rank per node and 48 OpenMP threads per MPI task.**

The scaling plot for the training of the CERN network on SNG as illustrated by the blue line compared to the linear scaling orange line. Figure 1 shows that the training exhibits almost linear scaling up to 256 nodes.

| Nodes | Training Time(S) per Epoch | Linear Time(S) per Epoch | Scaling Efficiency |
|---|---|---|---|
| 4 | 2302 | 2302 | - |
| 8 | 1238 | 1151 | 93% |
| 16 | 638 | 575.5 | 90.2% |
| 32 | 323 | 287.75 | 89.1% |
| 64 | 164 | 143.87 | 87.7% |
| 128 | 88 | 79.93 | 81.8% |
| 256 | 47 | 35.96 | 76.6% |
| 512 | 25 | 17.98 | 71.9% |

**Table 2: Shows the time taken in seconds for a single epoch of the CERN network to complete on SNG with 2 MPI ranks per node and hyperthreading with 48 OpenMP threads per MPI task.**

Table 2 shows the scaling results of the same training network with 2 MPI ranks per node and hyperthreading enabled, where each physical CPU core is split into two logical (virtual) cores. That are 1 MPI task per NUMA domain and 48 threads per MPI task. The standard configuration on SNG is each node has 2 NUMA domains, which means that each node has 2 MPI tasks and a total of 96 threads per node. Comparing the results of the 1 MPI task and 48 OpenMP threads per node with hyperthreaded configuration of 2 MPI tasks and 48 OpenMP threads per MPI task we noticed a performance increase in the training times of approximately 1.6 times.

| Nodes | Training Time(S) per Epoch | Linear Time(S) per Epoch | Scaling Efficiency |
|---|---|---|---|
| 4 | 959 | 959 | - |
| 8 | 507 | 479.5 | 94.6% |
| 16 | 264 | 239.75 | 90.8% |
| 32 | 137 | 119.87 | 87.5% |
| 64 | 72 | 59.93 | 83.3% |
| 128 | 39 | 29.96 | 76.8% |
| 256 | 21 | 14.98 | 71.4% |
| 512 | 12 | 7.49 | 62.5% |

**Table 3: Shows the time taken in seconds for a single epoch of the CERN network to complete with 4 MPI tasks per node and 12 OpenMP threads per MPI task on SNG.**

However, it is possible to configure SNG to allow clustering of the Skylake socket into a 4 NUMA domain and table 3 shows the scaling results of the same training network with the parameters of 4 MPI tasks per node and 12 threads per MPI task. This configuration results in a performance increase in the training times of a single epoch of approximately 2.38 times over the configuration with 2 MPI tasks per node and hyperthreading enabled.

Comparing the scaling plot in figure 1, for the 1 MPI task per node, against figure 2 configuration, with 4 MPI tasks per node, we noticed that the scaling efficiency is not as high as the 1 MPI task, but the time to solution is approximately 3.5 times faster for the configuration of 4 MPI tasks per node.

However, we encountered issues with interaction between the MPI libraries inside the container and the host MPI libraries when executing on more than 512 nodes, which often resulted in the application crashing due to MPI errors. Also, this instability in the MPI communication had a negative impact on the execution times of the epochs. SNG has a customized and system tuned Intel MPI version, which is the default and recommended MPI variant for the system. This version contains several bug fixes and optimizations that are not in the current version of Intel MPI, but will be introduced into future Intel MPI releases. OpenMPI is the only other MPI variant installed on the system, but we didn't use OpenMPI, as it is only recommended for experimental purposes and LRZ doesn't provide any functional or reliability guarantees.

Another issue we encountered was related to OmniPath; specifically, the Ubuntu container we were using had no support for psm2. That caused the application to crash with MPI errors. It was possible to resolve this issue by setting the fabric to TCP, but this had a negative impact on performance. Unfortunately, Intel doesn't provide OmniPath drivers for Ubuntu, which matches the custom version of the libraries installed on SNG. One option would be to use SLES/Leap or CentOS as the container OS, but typically data scientists prefer to use Ubuntu.

To resolve these issues, we installed a version of Intel MPI inside the container that corresponded to the host version to ensure that the 3DGAN network used the host MPI libraries during runtime. We then created a directory inside the container that corresponded to the host MPI library directory and we bind the host library directory into the container. This enables the host MPI libraries to be used during runtime. This method works, because Charliecloud imports the host environment settings.

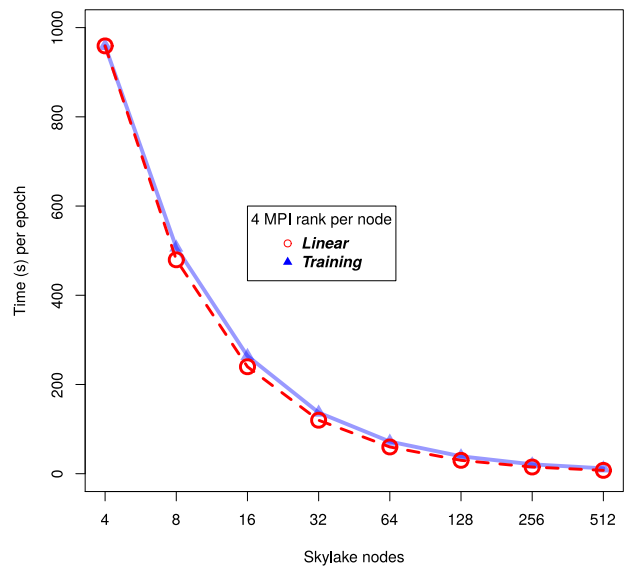

**Figure 2: Scaling plot for the training of the TensorFlow on SNG with 4 MPI ranks per node.**

| Nodes | Training Time(S) per Epoch | Linear Time(S) per Epoch | Scaling Efficiency |
|-------|------------------|------------------|-----------|
| 4 | 907.26 | 907.26 | - |
| 8 | 479.52 | 453.63 | 94.6% |
| 16 | 244.42 | 226.82 | 92.8% |
| 32 | 124.22 | 113.41 | 91.3% |
| 64 | 62.24 | 56.70 | 91.1% |
| 128 | 31.22 | 28.35 | 90..8% |
| 256 | 15.63 | 14.18 | 90.7% |
| 512 | 7.84 | 7.09 | 90.4% |
| 768 | 3.94 | 3.54 | 89.9% |

**Table 4: Shows the time taken in seconds for a single epoch of the CERN network to complete with 4 MPI tasks per node and 12 OpenMP threads per MPI task using the host Intel MPI libraries on SNG.**

Table 4 shows the scaling results of the same 3DGAN training network but using the host MPI libraries rather than the standard MPICH libraries inside the container. This modification greatly increased the stability of the MPI communication and enabled the network to be successfully trained on 768 nodes, which is a single island of SNG. In addition to the increase in stability, the scaling has improved to approximately 90% efficiency, which is what we would expect for the distributed training of the 3DGAN model.

## 5.2 Multi-Node Scaling Non-Containerized

In this section, we discuss the results of executing the 3DGAN network from CERN on HPC clusters, which did not use containerized solutions for deployment.

Table 5 shows the time taken to complete a single epoch on a 64 node cluster at Intel labs. Where each node contains 40 core Intel Xeon Gold 6148 Skylake processors. We modified the MPI and OpenMP parameters to 4 MPI tasks per node and 10 threads per MPI task. So that it would conform to the configuration of the node.

| Nodes | Training Time(S) per Epoch | Linear Time(S) per Epoch | Scaling Efficiency |
|-------|------------------|------------------|-----------|
| 1 | 7453 | 7453 | - |
| 2 | 3797 | 3726.5 | 98.14% |
| 4 | 1934 | 1863.25 | 96.34% |
| 8 | 990 | 931.63 | 94.1% |
| 16 | 504 | 465.81 | 92.42% |
| 32 | 263 | 232.91 | 88.55% |
| 64 | 132 | 116.45 | 88.22% |

**Table 5: Shows the time taken in seconds for a single epoch of the CERN network to complete with 4 MPI tasks per node**

**and 10 OpenMP threads per MPI task on an HPC cluster at Intel.**

The results shown in table 5 for the non-containerized version of TensorFlow framework shows an increase in performance of approximately 20% over the 48 core Platinum 8174 processors with 2 MPI tasks per core and 48 threads per task on SNG as shown in table 2. We would have expected that the performance would be better on SNG as the processors are more performant. We suspected that number of MPI tasks per node was responsible for this performance difference. So, to verify this hypothesis we executed the training again on SNG with 4 MPI tasks per node. The result for training time for 1 epoch on SNG was approximately 1.92 times faster than on the cluster at Intel labs. This large improvement in performance of the training on SNG compared to the Intel cluster is because SNG has more processor cores per node, which run at a higher CPU clock frequency and the network fabric & communication libraries have been optimized.

| Nodes | Training Time(S) per Epoch | Linear Time(S) per Epoch | Scaling Efficiency |
|-------|------------------|------------------|-----------|
| 1 | 17831 | 17831 | - |
| 2 | 8998 | 8915.5 | 99.1% |
| 4 | 4545 | 4457.75 | 98.08% |
| 8 | 2288 | 2228.87 | 97.4% |
| 16 | 1151 | 1114.44 | 96.8% |
| 32 | 581 | 557.22 | 95.9% |
| 64 | 293 | 278.61 | 95.1% |
| 128 | 148 | 139.60 | 94.1% |

**Table 6: Shows the time taken in seconds for a single epoch of the CERN network to complete with 4 MPI tasks per node and 11 threads per MPI task on the Stampede2 cluster at TACC.**

Table 6 shows the training times for 1 epoch on the Stampede2 HPC cluster at TACC, with each node containing 2 sockets and 24 cores per socket Intel® Xeon® 8160 CPUs. Notice that the training times are significantly worse than on both the Intel lab cluster and SNG. The training time for 1 epoch on the Intel lab cluster was approximately 2.28 times faster than on Stampede2. While the training time for 1 epoch on SNG was approximately 4.29 times faster than on Stampede2. The training of the network done at the Texas Advanced Computing Center (TACC) was with older versions of MKL-DNN, TensorFlow and Keras and as we ran out of compute credits, we were unable to use recent versions of MKL-DNN, Keras, Horovod and TensorFlow.

When comparing the performance of the 3DGAN training network when executed from within a Charliecloud container with the TensorFlow framework built directly on the system, we don't see a degradation in performance. Also, the results from Stampede2 indicates that a significant amount of the performance is due to improvements in the numerical and parallel distribution libraries such as MKL-DNN and Horovod with MLSL. One reason for this improvement is that MKL-DNN has been optimized for mathematical operators used in deep neural networks, as well as optimizations for operations on very small matrices. This Hypothesis is supported by the comparison of measured inferencing results for TensorFlow versions 1.4 and 1.9 on Stampede2 by Vallecorsa et al. [28], Intel Python distribution benchmarks [29] and SIMD optimization techniques demonstrated by Georganas et al. [30] which have been incorporated into Intel's MKL-DNN.

## 5.3 Compute Performance

In this section we evaluate the compute performance of the convolution kernel optimized with the Intel MKL-DNN libraries used in the training of the 3DGAN network.

In our measurements we rerun the same setup as we had done previously with the scaling tests, but also measure the FLOP rate and calculate the percentage of the theoretical maximum performance for a single convolution kernel. The theoretical peak performance of the SNG system is 26.9 PFLOPS and for a single node is 4.15 TFLOPS.

However, the actual production mode processor operating frequency is 2.3GHz rather than 2.7GHz which is the theoretical maxim CPU frequency used in the systems HPL benchmark [31]. Also, we are assuming that the throughput of the 2 AVX units per core is a 512 bit FMA operation per cycle. Also, in "real world" execution mode the frequency of the AVX units are less than that of the CPU in normal operational mode. So, we expect there will be a big difference between the theoretical peak performance and the actual maximum performance of the compute node in production mode. Ideally, we would like to compare the performance against HPL results for the system in production mode and we would expect the relative performance to be much better. The HPL performance with the non-production setting was 19.4766 petaflops and the theoretical peak is 26.8739 petaflops. However, due to costs, HPL runs are only performed during the bring up phase of the system to provide data for the HPC Top500 list.

| Nodes | Measured Performance petaflops | Percentage of Theoretical Peak |
|---|---|---|
| 4 | 0.01099 | 66.17% |

| 8 | 0.02199 | 66.21% |
| 16 | 0.04450 | 67.01% |
| 32 | 0.08386 | 63.14% |
| 64 | 0.17313 | 65.17% |
| 128 | 0.31878 | 67.60% |
| 256 | 0.70547 | 66.39% |
| 512 | 1.39412 | 65.60% |
| 768 | 2.08143 | 65.29% |

**Table 7: Shows the compute performance of a single convolution kernel in petaflops.**

The results shown in table 7 for the compute performance of a single convolution kernel indicate that approximately 66% of the theoretical peak compute performance is achieved on SNG. This performance is extremely good, especially when you consider that in production mode the actual peak performance is significantly less than that of the theoretical max performance for the compute nodes.

## 5.4 Overheads

HPC systems come with limited amount of resources, such as system memory and storage. It is important that new software stacks, such as the Charliecloud, exhibit minimal overhead in terms of runtime and utilization of the system resources. For example, the useable amount of system memory on a single compute node on SNG is limited to approximately 77 GB and exceeding it will cause the job and compute node to crash. In table 8, we observe negligible performance overhead for TensorFlow version 1.9 workloads on a 4 node Xeon gold 6148, 2 socket, 20 cores, 2 threads per core Skylake cluster using Charliecloud for executing topologies such as ResNet-50 and AlexNet.

| Benchmark | TF Throughput with Charliecloud (img/s) | TF Throughput without Charliecloud (img/s) |
|---|---|---|
| AlexNet with cifar10 | 1968 | 1973 |
| ResNet-50 | 75 | 74 |

**Table 8: Achieved throughput (img/s) for AlexNet and ResNet-50 based on TensorFlow 1.11 with and without Charliecloud.**

Memory overhead measurements were negligible when running the workloads through the Charliecloud runtime environment compared to running directly on the host system as illustrated in table 9.

| Benchmark | Free System Memory with Charliecloud (GB) | Free System Memory without Charliecloud (GB) |
|---|---|---|

| AlexNet with cifar | 331.29 | 331.33 |
| ResNet50 with imagenet | 324.47 | 324.89 |

**Table 9: Free system memory for AlexNet and Resnet50 based on TensorFlow 1.11 with and without Charliecloud.**

A.Torrez et al. [32] describe the performance impacts of HPC container runtimes on HPC benchmarks such as High Performance LINPACK (HPL), High Performance Conjugate Gradient (HPCG) and STREAM.

# 6 Conclusions

We were able to set-up the software stack, implement and test the full experimental pipeline on up to 512 nodes using a configuration of one MPI task and 48 OpenMP threads per node in less than one day by using an Intel optimized version of TensorFlow, MPICH variant of MPI inside the container and Intel MPI on the HPC system. Other results were collected over a longer time period specifically the results with different MPI & OpenMP configurations and node configurations

The results are extremely promising and justify the effort we have made in developing a secure containerization solution for deploying AI on HPC systems.

Also, we have shown that it is possible to not only scale the distributed training of the neural network to 100's of nodes with an extremely good efficiency. We were also able to achieve a significant compute performance for the convolution kernel. However, to achieve this performance on the very large HPC systems we have to use hardware vendor specific numerical libraries. In addition to enable the training to be distributed across 100's or 1000's of compute nodes we have to use the same MPI libraries inside the container and on the host system. This means that performance portability of containerized applications on large systems is not possible due to the custom software and system configurations; such as optimized MPI, network fabric and filesystem. However, it is possible to quickly create containers with specific HPC system optimizations by installing the vendors implementation of MPI and network fabric drivers inside the containers.

# 7 Future Work

This work is to be intended as proof of concept for deploying AI and ML software at petascale on a secure HPC system, and we plan to provide a list of best practices and recipes to share for production-ready deployments on large HPC systems. On the technical side, we would also like to extend this work to investigate the deployment of other AI and ML frameworks, and validate the procedure on new workloads and existing HPC applications.

## ACKNOWLEDGMENTS

The authors gratefully acknowledge the Gauss Centre for Supercomputing e.V. (*www.gauss-centre.eu*) for funding this project by providing computing time on the GCS Supercomputer SuperMUC at Leibniz Supercomputing Centre (*www.lrz.de*). The authors also acknowledge support by the German Federal Ministry of Education and Research via the project DeToL (01IS18046).